\documentclass[conference]{IEEEtran}

\usepackage[T1]{fontenc}
\usepackage[utf8]{inputenc}
\usepackage{bm}
\usepackage[caption=false]{subfig}
\usepackage{tikz}
\usepackage{pgfplots}
\usepackage{amsmath}
\usepackage{dsfont}
\usepackage{amssymb}
\usepackage{graphicx}
\usepackage{cite}
\usepackage{fst}
\usepackage{tumcolor}
\usepackage{siunitx}
\usepackage{algorithm}
\usepackage{algpseudocode}
\usepackage{booktabs}
\usepackage{textcomp}
\usepackage[normalem]{ulem}
\usepackage{csquotes}
\usepackage{multirow}
\pgfplotsset{compat=newest}

\algnewcommand\algorithmicinput{\textbf{INPUT:}}
\algnewcommand\INPUT{\item[\algorithmicinput]}

\newcommand{\SNR}{\text{SNR}}

\DeclareSIUnit\bpcu{bpcu}
\DeclareSIUnit\bpQs{bpQs}

\usetikzlibrary{external}
\IEEEoverridecommandlockouts    
    
\begin{document}

\title{Comparison of Geometric and Probabilistic Shaping with Application to ATSC 3.0}

\author{
\IEEEauthorblockN{Fabian Steiner\IEEEauthorrefmark{1}, Georg Böcherer\IEEEauthorrefmark{2}}
\IEEEauthorblockA{\IEEEauthorrefmark{1}Fachgebiet Methoden der Signalverarbeitung\\ \IEEEauthorrefmark{2}Institute for Communications Engineering\\Technical University of Munich\\Email: \{fabian.steiner, georg.boecherer\}@tum.de}

\thanks{F. Steiner was supported by the TUM--Institute for Advanced Study, funded by the German Excellence Initiative and the European Union Seventh Framework Program under grant agreement n\textdegree{ }291763. Georg Böcherer was supported by the German Federal Ministry of Education and Research in the framework of an Alexander von Humboldt Professorship.}

}

\markboth{}{}%

\maketitle

\begin{abstract}
 In this work, geometric shaping (GS) and probabilistic shaping (PS) for the AWGN channel is reviewed. Both
 approaches are investigated in terms of symbol-metric decoding (SMD) and bit-metric decoding (BMD). For
 GS, an optimization algorithm based on differential evolution is formulated. Achievable rate analysis reveals 
 that GS suffers from a \SI{0.4}{dB} performance degradation compared to PS when BMD is used. 
 Forward-error correction simulations of the ATSC~3.0 modulation and coding formats (modcods) confirm the theoretical findings. In
 particular, PS enables seamless rate adaptation with one single modcod and it outperforms ATSC 3.0 GS modcods by
 more than \SI{0.5}{dB} for spectral efficiencies larger than \num{3.2} bits per channel use.
\end{abstract}

\section{Introduction}

Bandwidth-efficient communication requires to use higher-order modulation formats, such as $M$-\ac{ASK}, $M$-\ac{QAM}
or $M$-\ac{APSK} constellations. On the \ac{AWGN} channel model, discrete, equidistant constellations with 
uniform signaling result in a gap to capacity of \SI{1.53}{dB} in the high \ac{SNR} 
regime~\cite[Sec.~IV-B]{forney_efficient_1984}.

In order to compensate for this performance loss, \ac{PS} and \ac{GS} can be employed in
order to mimic a \enquote{Gaussian-like} shape of the constellation. 
\ac{PS} imposes a non-uniform distribution on a set of equidistant constellation points. We refer to \cite[Sect.~II]{bocherer_bandwidth_2015}
for a literature review on probabilistic shaping approaches. As the desired distribution needs to be ensured at the 
channel input, some schemes perform the shaping operation \emph{after} \ac{FEC} encoding and it is reversed before decoding.
This is prone to error propagation and usually requires iterative processing at the receiver~\cite{valenti_constellation_2012}. In \cite{bocherer_bandwidth_2015},
\ac{PAS} is introduced which concatenates encoding and shaping in reverse and enables
a simple receiver setup. 

Geometric shaping employs a uniform distribution on non-equidistant constellation
points. The authors of \cite{sun_approaching_1993} show that this approach achieves the
capacity of the \ac{AWGN} channel if the number of constellation points goes to infinity. In
\cite{barsoum_constellation_2007}, the effect of geometric shaping is investigated in terms 
of the achievable rate when both \ac{SMD} and \ac{BMD} are employed on one-dimensional 
constellations. The numerical results indicate that both optimization criteria lead to 
different optimized constellations. Recently, geometrically shaped constellations were included in the
DVB-NGH~\cite{etsi2013dvbngh,gomez-barquero_dvb-ngh:_2014} and ATSC~3.0 standards~\cite{atsc30,loghin_non-uniform_2016},
where they are generally referred to as \acp{NUC}.

The contributions of the present work are twofold. 
\begin{itemize}
 \item We provide a comprehensive comparison
of both \ac{PS} and \ac{GS} in terms of their information theoretic achievable rates for \ac{SMD} and \ac{BMD}. To this end,
we propose a \ac{DE}~\cite{storn_differential_1997} based optimization approach to obtain optimized 
geometrically shaped constellations. The results show that \ac{GS} has a gap to capacity of about
\SI{0.4}{dB}, when \ac{BMD} is used. In contrast, \ac{PS} with \ac{BMD} virtually achieves capacity.
\item We compare a selection of ATSC~3.0 modcods, i.e., modulation order and code rate combinations, to
a \ac{PS} system operating with a single modcod using \ac{PAS}~\cite{bocherer_bandwidth_2015}. \ac{FEC} simulations show that
the information theoretic gains translate into practical, coded performance improvements to a full extent.
\end{itemize}

The work is organized as follows. In Sec.~\ref{sec:system_model}, we present the 
system model. Sec.~\ref{sec:shaping} states the optimization procedures
for both \ac{GS} and \ac{PS}. We present \ac{FEC} simulation results in
Sec.~\ref{sec:coded_performance} and conclude in Sec.~\ref{sec:conclusion}.

\section{System Model}
\label{sec:system_model}

We consider the discrete time \ac{AWGN} channel
\begin{equation}
 Y_i = X_i + Z_i, \quad i = 1, \ldots, n_\tc\label{eq:system_model}
\end{equation}
for a transmission block of $n_\tc$ channel uses. The time indices are omitted in the following. As we consider both one and two dimensional constellations,
the noise $Z$ is either zero mean Gaussian, or zero-mean, circular symmetric complex Gaussian. Accordingly, the channel input $X$ originates from an $M$-ary real or
complex signaling set $\cX$.  We define $\SNR=\E{\abs{X}^2}/\E{\abs{Z}^2}$. The random variables $X$ and $Z$ are either both real or both complex; the resulting \ac{SNR} is
the same in both cases. For each dimension, the capacity is given by
\begin{equation}
 \sfC(\SNR) = \frac{1}{2}\log_2(1+\SNR).
\end{equation}

For practical systems, a Gaussian codebook is not feasible so that discrete constellations like 
\ac{ASK} and \ac{QAM} are employed. 
To combine $M=2^m$-ary higher-order constellations
with \ac{FEC}, a code over $\text{GF}(M)$ can be used and the field elements
are directly mapped to constellation points. However, this usually comes at the
price of increased decoding complexity~\cite[Sec. IV]{bennatan_design_2006}.
At the receiver, different decoding metrics can be employed. For non-binary \ac{FEC}, one typically uses
\ac{SMD} with the decoding metric $q(x,y) = p_{Y|X}(y|x)P_X(x)$ and an achievable rate is given by the mutual 
information $\I(X;Y)$, i.e.,
\begin{equation}
 \Rsmd(P_X,\SNR) = \I(X;Y)\label{eq:r_smd}.
\end{equation}

To use a binary \ac{FEC}, a binary labeling of the signal points has to be introduced. 
Hence, each signal point $x_{\vb} \in\cX$ is assigned a binary $m$-bit 
label of the form $\vb = b_1b_2\ldots b_m$ with $b_i \in\{0,1\}, i = 1,\ldots,m$.
In this case, usually a \emph{pragmatic} approach with a bitwise-metric is pursued, consisting of a binary soft demapper followed by a binary
decoder. This approach was introduced in \cite{zehavi_8-psk_1992} and is now often called \ac{BICM}~\cite{caire_bit-interleaved_1998}.
It can be motivated by a mismatched decoding perspective~\cite{martinez_mismatched}, where
the decoding metric $q(x,y) = q(x_{\vb},y) = \prod_{i=1}^m p_{y|B_i}(y|b_i)P_{B_i}(b_i)$ is used
to arrive at the information theoretic model of $m$ parallel channels with
\[
 p_{Y|B_i}(y|b) = \frac{1}{P_{B_i}(b)} \sum_{\xi\in\cX_i^b} p_{Y|X}(y|\xi) P_X(\xi).
\]
Here, the set $\cX_i^b\subseteq\cX$ refers to all constellation points which have the $i$-th bit in their binary label set to $b$.
In~\cite{bocherer2014achievable}, it was recently shown that an
achievable rate is given by
\begin{equation}
 \Rbmd(P_X,\SNR) = \left[\entr(\vB) - \sum_{i=1}^m \entr(B_i|Y_i)\right]^+. \label{eq:r_bmd}
\end{equation}

\begin{myremark}
 In case of uniform inputs, the \ac{BMD} achievable rate~\eqref{eq:r_bmd} can be rewritten as $\Rbmd = \sum_{i=1}^m\I(B_i;Y)$, which is
 commonly known as \emph{BICM capacity} \cite{martinez_mismatched}.
\end{myremark}

We will also employ the notation $\Rbmd^{-1}(P_X,R)$ to denote the required $\SNR$ to achieve a \ac{BMD}
rate $R$, i.e., $\Rbmd(P_X,\SNR) = R$. If $P_X$ is omitted, a uniform distribution
on the constellation points is assumed in both cases. 

Using the chain rule of mutual information, it can be shown that $\Rbmd \leq \Rsmd$ 
and clearly, $\Rsmd \leq \sfC$. The rate $\Rbmd$ depends on the employed 
labeling and a \ac{BRGC}~\cite{gray1953pulse} usually performs well.

\section{Geometric and Probabilistic Shaping}
\label{sec:shaping}

\subsection{Geometric Shaping}
\label{sec:geometric_shaping}

Geometric constellation shaping employs non-equidistant spacing of the constellation points
with a uniform distribution. The best constellation
depends on the $\SNR$ and the employed metric. To facilitate the solution approach in the following,
we refine \eqref{eq:system_model} w.l.o.g. and impose $\E{\abs{X}^2} \leq 1$
such that the SNR is given by $1/\E{\abs{Z}^2}$. The optimization
problem to find the optimized constellation $\cX^*$ can be formulated as
\begin{equation}
 \cX^* = \argmax_{\substack{\cX: \E{\abs{X}^2} \leq 1\\\abs{\cX} = M}} \sfR_{\{\tbmd,\tsmd\}}(\SNR).\label{eq:opt_geom}
\end{equation}
For both metrics, the optimization in $\cX$ is non-convex.
The works~\cite{barsoum_constellation_2007,zoellner_optimization_2013} employed 
\enquote{constrained non-linear optimization algorithms} without providing more
details on the actually employed optimization procedure. In~\cite{kayhan_constellation_2012},
simulated annealing is used to optimize \ac{APSK} constellations.
Initial investigations by using standard, black box interior point
algorithms like Matlab's \texttt{fmincon} showed that the obtained optimization results
depend on the initialization, which suggests that only locally optimal
solutions are found.

In the following, we propose an optimization based on differential evolution~\cite{storn_differential_1997},
which is a genetic algorithm and appears to find the global optimum, i.e., \ac{DE}
recovered previously reported results from arbitrary valid starting points.

\begin{algorithm}[t]
\begin{algorithmic}[1]
\INPUT $\SNR$, constellation size $M$, candidate set size $P$, number of generations $G$,
crossover probability $p_c$, amplification factor $F$.
\State Create feasible initial population set $\{\tilde\cX_p\}_{p=1}^P$ at random.\label{alg:initial_pop}
\State Evaluate metric $\sfR_{\{\tbmd,\tsmd\}}$ for each population member.
\For{$g = 1, \ldots, G$}
  \For{$p = 1, \ldots, P$}
    \State Choose $r_1\neq r_2\neq r_3$ randomly from $\{1,\ldots,P\}$.
    \State $\vT = \texttt{map}(\tilde\cX^{(g-1)}_{r_1} + F\cdot(\tilde\cX^{(g-1)}_{r_2} - \tilde\cX^{(g-1)}_{r_3}))$\label{alg:get_new}
    \State $\vT = \texttt{mutate}(\vT,\tilde\cX^{(g-1)}_p,p_\tc)$
    \State Evaluate metric of new candidate $\vT$. 
    \State Set $\tilde\cX^{(g)}_p = \vT$, if metric has improved. 
  \EndFor
  \If{all population members have the same metric}
  \State Stop.
  \EndIf
\EndFor
\end{algorithmic}
\caption{Summary of the \ac{DE} algorithm to find the best constellation for a given SNR.}
\label{alg:DE_alg}
\end{algorithm}

\ac{DE} starts with an initial population $\{\tilde\cX_p\}_{p=1}^P$ of candidate constellations 
(see line~\ref{alg:initial_pop}). In each generation, a population member experiences a mutation. For this, \ac{DE}
randomly selects three distinct population members and combines them as shown in line~\ref{alg:get_new}. As
the result of this operation may violate the feasible set, the function $\texttt{map}(\cdot)$
implements a bounce back strategy. Eventually, the new candidate constellation is generated by replacing each component of $\tilde\cX_p^{(g-1)}$
with probability $p_\tc$ by the corresponding entry of $\vT$. If the metric for the new candidate $\vT$ has improved
we keep it, otherwise we set $\tilde\cX_p^{(g)} = \tilde\cX_p^{(g-1)}$. We stop after $G$ generations or once all population
members have the same objective function value.

We distinguish between one-dimensional \ac{GS} (1D-GS) and two-dimensional \ac{GS} (2D-GS)
and exploit symmetry to decrease the number of optimization parameters.

For 1D-GS and an $M$-ary 1D constellation (1D-GS 1D-NUC), each of the $M/2$ components of $\tilde\cX_p$ 
is constrained to the non-negative real axis and the augmented, final constellation $\cX_p$
with the negative part must fulfill the power constraint. A two-dimensional 1D-GS $M$-ary constellation (1D-GS 2D-NUC)
can be obtained by the Cartesian product of two copies of 1D-GS $\sqrt{M}$-ary 1D-\acp{NUC}. 

For 2D-GS, the population members are restricted to the first quadrant of the complex plane and
$(M/4) \cdot 2$ real variables have to be optimized ($M/4$ for the real and $M/4$ for the imaginary part).
This introduces additional \acp{DOF} and leads to larger achievable rates.

To remain in the feasible set, i.e., on the real non-negative axis for 1D-GS and the first quadrant for 2D-GS, 
the \texttt{map} function (see line~\ref{alg:get_new}) replaces any negative real or imaginary part by
its absolute value and rescales it to meet the power constraint.

In our trials, we used an amplification factor $F=\num{0.5}$ and a crossover probability $p_\tc = 0.88$. The number of generations is set to 
$G = \num{10000}$ and the population size was chosen depending on the number of \acp{DOF} as $P = 5\cdot\text{DOF}$. 
Choosing this parameter thoroughly turned out to be crucial in our experiments: Setting it too small, the optimum may 
not be found and setting it too large, the number of generations would not suffice. In the low
and medium SNR regime, usually 100 to 1000 generations are enough to observe convergence.

If the metric targets \ac{BMD} rates, the influence of the labels has to be taken into account as well. For 1D-GS, we
randomly assign each component of $\tilde\cX_p$ a $\log_2(M)-1$ bit label. The labels for the augmented constellation $\cX_p$ are then obtained
by first replicating and then prefixing each half with a zero and one, respectively. For 2D-GS, the same approach applies, however, each quadrant
in the augmented constellation is prefixed by one of the four two-bit labels 00, 10, 11 and 01 in an ordered manner, which
is consistent with ATSC~3.0.

\subsection{Probabilistic Shaping}
\label{sec:pas}

In contrast to \ac{GS}, \ac{PS} uses a non-uniform distribution over equidistant constellation
points. Contrary to the \ac{GS} case, we instantiate \eqref{eq:system_model} for the real case, i.e., w.l.o.g. $Z\sim\cN(0,1)$
and set $X = \Delta \tilde X$, where the positive, real valued parameter denotes the spacing between the 
constellation points and $\tilde X \in \{\pm 1, \pm 3, \ldots, \pm 2^{m-1}\}$. The 
corresponding optimization problem reads as
\begin{equation}
 P_X^* = \argmax_{P_X, \Delta > 0: \E{\abs{\Delta \tilde X}^2} \leq \SNR} \sfR_{\{\tsmd,\tbmd\}}(P_X,\SNR)\label{eq:opt_pas}.
\end{equation}

For fixed $\Delta$, $\Rsmd$ is convex in $P_X$~\cite[Ch.~4.4]{gallager1968} and the following nested
approach can be taken:
\begin{enumerate}
 \item Parameter $\Delta$ is optimized using a simple line search.
 \item For each fixed $\Delta$ during the line search, perform the optimization in $P_X$, using efficient and globally optimal optimization algorithms like
 Blahut-Arimoto~\cite{blahut_capacity} or Cutting-Plane based approaches~\cite{huang_characterization_2005}.
\end{enumerate}
For \ac{BMD}, the sequential approach can still be used, but $R_\tbmd$ is not convex in $P_X$ anymore.

This problem can be circumvented by optimizing over the \ac{MB} distribution family~\cite{kschischang_pasupathy_maxwell}.
As the numerical evaluations in Sec.~\ref{sec:achievable_rates_cmp} show, even this suboptimal approach virtually achieves 
capacity and outperforms the \ac{GS} results, which are conjectured to be globally optimal. A detailed comparison of the 
optimal and \ac{MB} based performance for both \ac{SMD} and \ac{BMD} is provided in \cite[Table~I, III]{bocherer_bandwidth_2015}. 

Two-dimensional \ac{PS} is obtained by taking two copies of the one-dimensional \ac{PS} constellations.

\subsection{Achievable Rates Comparisons}
\label{sec:achievable_rates_cmp}

In the following, we compare the \ac{BMD} and \ac{SMD} achievable rates for both geometrically and probabilistically shaped constellations. As a performance
metric we resort to the $\SNR$ gap to capacity. It is defined as
\begin{equation}
 \Delta\SNR = \SNR - \sfC^{-1}(\Rbmd(P_X^*,\SNR))\label{eq:snr_gap},
\end{equation}
where $\sfC^{-1}(\cdot)$ represents the inverse capacity functional of the real or complex \ac{AWGN} channel, respectively. $P_X^*$ is either the uniform 
distribution $1/\abs{\cX^*}$ on $\cX^*$ of \eqref{eq:opt_geom} for \ac{GS} or the optimal distribution for \ac{PS} as the solution of \eqref{eq:opt_pas}.
The obtained constellations are available from \cite{website_scc2016}.

\begin{figure}[h]
 \footnotesize
 \tikzsetnextfilename{rates_1D_cm}
 \includegraphics{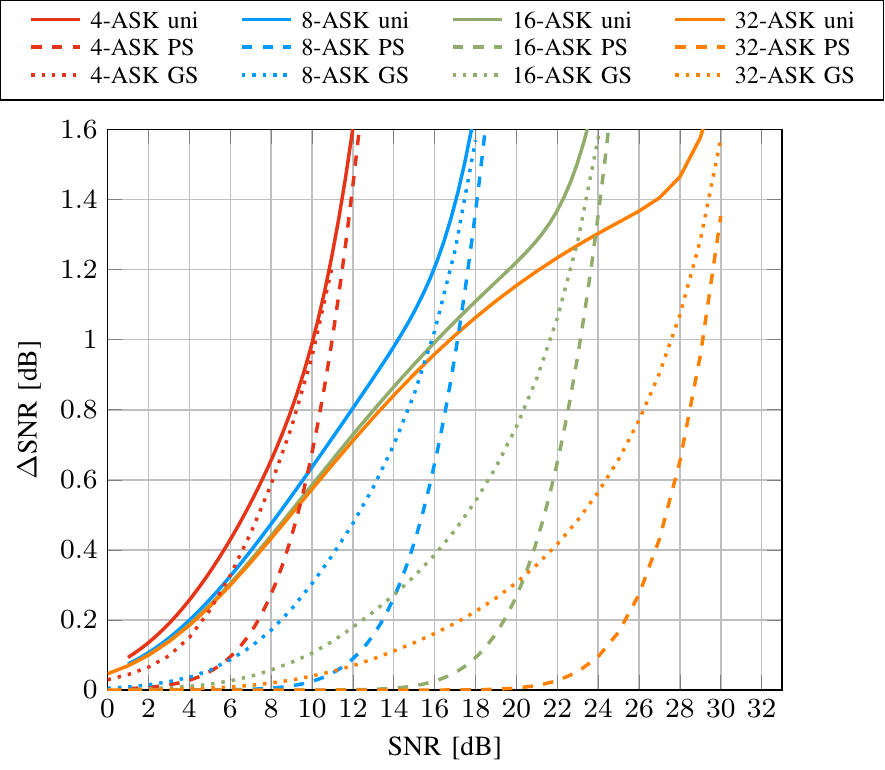}
 \vspace{-1\baselineskip}
 \caption{\ac{SMD} gap to capacity in \si{dB} for 1D-GS and \ac{PS}.}
 \label{fig:1D_rates_cm}
\end{figure}

\begin{figure}[h]
 \footnotesize
 \tikzsetnextfilename{rates_1D}
 \includegraphics{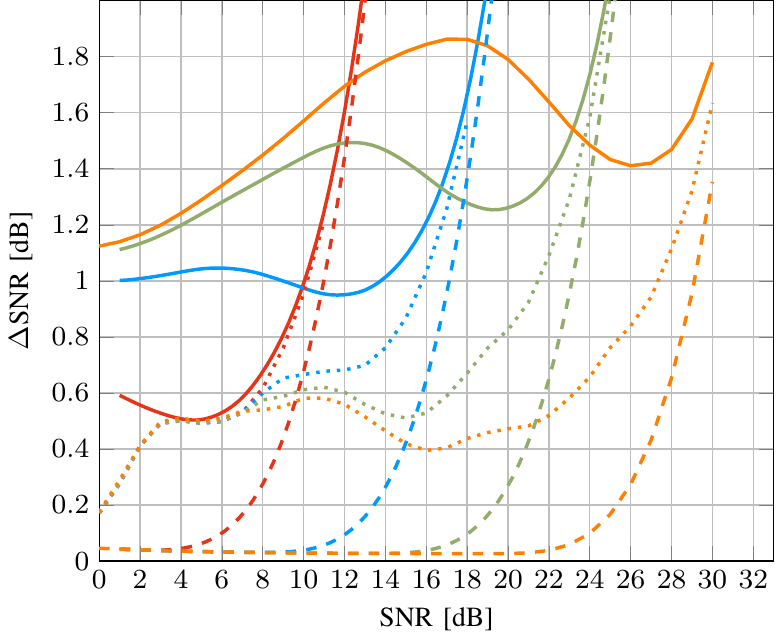}
 \vspace{-1\baselineskip}
 \caption{\ac{BMD} gap to capacity in \si{dB} for 1D-GS and \ac{PS}.}
 \label{fig:1D_rates}
\end{figure}

Fig.~\ref{fig:1D_rates_cm} illustrates the gap to capacity for the optimized one-dimensional $\{4, 8, 16, 32\}$-ASK constellations in case of \ac{SMD}. As a reference, we also
plot the gaps for uniform, equidistant constellations. As derived in previous work~\cite{sun_approaching_1993}, the shaping gain of geometrically
shaped constellations increases with the constellation size. 

Fig.~\ref{fig:1D_rates} provides the equivalent evaluation for \ac{BMD}. Here, the gap to capacity does not exhibit a monotonous behavior and is particularly larger
in the low to medium \ac{SNR} regime, as there is an additional \ac{BMD} penalty. \ac{PS} shows improved performance over the whole range of constellations and \ac{SNR} 
values. In particular, we observe that the gap to capacity remains almost constant in the order of \SI{e-2}{dB} (with an improvement of more than \SI{0.4}{dB} compared \ac{GS})
and virtually vanishes for \ac{SMD}, which emphasizes its applicability to both \ac{BMD} and \ac{SMD} receiver architectures.

\begin{figure}[h]
 \footnotesize
 \tikzsetnextfilename{rates_2D}
 \includegraphics{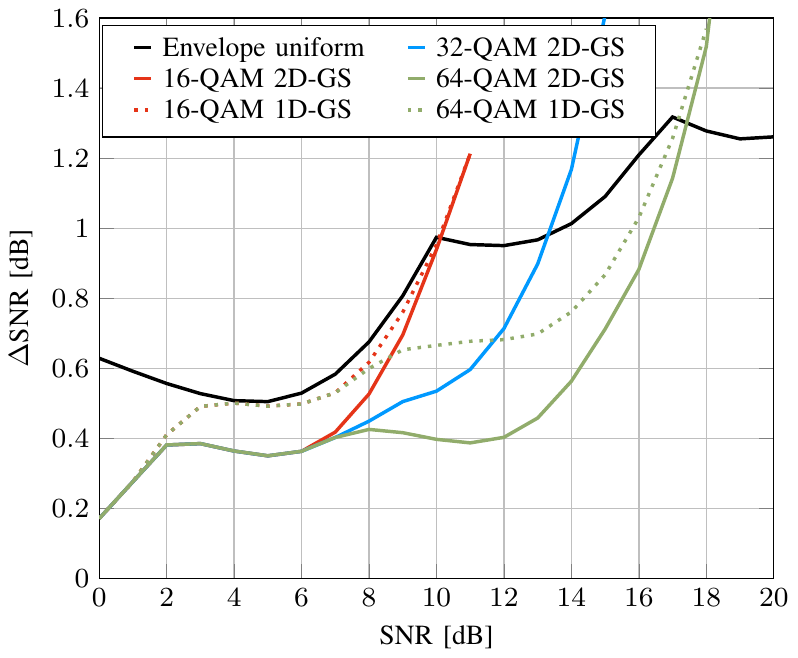}
 \vspace{-1\baselineskip}
 \caption{\ac{BMD} gap to capacity in \si{dB} for 2D-GS.}
 \label{fig:2D_rates}
\end{figure}

In Fig.~\ref{fig:2D_rates}, the gap to capacity for 1D-GS $\{16,64\}$-ary 2D-\acp{NUC} and 2D-GS $\{16, 32, 64\}$-ary constellations are shown. 
The benefits of the additional degrees of freedom are clearly visible.

Summarizing the above results, it becomes evident that both approaches are able to improve the \ac{SE} of a communication system
compared to uniform, equidistant signaling. \ac{PS} is clearly advantageous for the considered constellation sizes for both \ac{BMD} and 
\ac{SMD}. Hence, the statement of~\cite{barsoum_constellation_2007}, saying
that ``any gain in capacity which can be found via probabilistic shaping can also be achieved or exceeded
solely through geometric shaping'' should be considered with appropriate caution, as it implicitely assumes a much larger
constellation size for \ac{GS}. We illustrate this with two examples for \ac{SMD}.

\begin{example}
 In~\cite{sun_approaching_1993}, the authors provide a signal set construction, where the constellation points are chosen as the
 centroids of equiprobable quantiles of the Gaussian distribution and show that it is capacity-achieving for $M\to\infty$. 
 An equidistant 8-ASK constellation with optimized distribution using the procedure of Sect.~\ref{sec:pas} for \SI{10}{dB} yields  
 $\Rsmd(P_X^*,\SI{10}{dB}) = \SI{1.726}{\bpcu}$. To achieve the same rate, a number of $M = 50$ constellation points have to be 
 used for their \ac{GS} approach.
\end{example}

\begin{example}
In \cite{meric_approaching_2015}, the author describes a practical scheme to construct capacity-approaching 
\ac{APSK} constellations consisting of $n$ rings with $n$ constellation points, each. We consider the same case
as in Example~1. The \ac{PS} rate gap, i.e., $\sfC(\SI{10}{dB}) - \Rsmd(P_X^*,\SI{10}{dB})$, equals \num{0.0037} bits per
real dimension. According to \cite[Fig. 2b]{meric_approaching_2015}, this requires an \ac{APSK} constellation
with much more than $35^2=1225$ points and additional two-dimensional demapping.
\end{example}

Similar observations can be found in \cite{wu_impact_2010}, where the authors investigate the impact of constellation
cardinality on the effect of approaching the \ac{AWGN} channel capacity. They show that the convergence speed of methods like
\cite{sun_approaching_1993} is only $\cO(1/M^2)$ (and thus requiring large constellation sizes), whereas the use of Gauss
quadratures that involves both geometrical and probabilistic shaping is able to approach capacity exponentially fast in the
constellation size.

\section{\ac{FEC} Simulation Results}
\label{sec:coded_performance}

\subsection{Combining of \ac{GS} and \ac{PS} with \ac{FEC}}

We consider soft-decision based binary-input \ac{FEC} schemes in the following, where the decoder uses real valued
\acp{LLR}
\begin{equation}
 L_i = \log\left(\frac{p_{Y|B_i}(y|0)}{p_{Y|B_i}(y|1)}\right) + \log\left(\frac{P_{B_i}(0)}{P_{B_i}(1)}\right), \quad i = 1,\ldots, m.\label{eq:llr}
\end{equation}
The first term constitutes the channel log-likelihood, whereas the second term represents the priors and evaluates to zero for
uniform input distributions.

For \ac{GS}, the combination with \ac{FEC} is straightforward and does not require any modifications. However, it remains to note that 2D-GS 2D-\acp{NUC} 
require two-dimensional demapping, so that the \ac{LLR} calculation has increased complexity.

\ac{PAS}~\cite[Sec.~IV]{bocherer_bandwidth_2015} employs a \emph{reverse concatenation} of \ac{FEC} encoding and shaping, while exploiting
the symmetry of the optimal input distribution and systematic encoding:
\begin{itemize}
 \item The symmetric input distribution factors into independent random variables representing the amplitude $A$ and sign $S$, i.e., $P_X(x) = P_A(\abs{x}) P_S(\sign(x))$.
 While $P_A$ is non-uniform on $\{1,3,\ldots,2^m-1\}$, $P_S$ is uniform on $\{-1,+1\}$.
 \item The binary representation of the amplitude values is systematically encoded and copied to the codeword, so that their distribution is preserved. The calculated parity-check bits 
 are approximately uniform and can be used for the sign part.
\end{itemize}
At the receiver side, the decoder is made aware of the shaping with the help of the 
priors so that no additional deshaping operation has to be performed. The generation of non-uniform amplitudes from uniformly distributed information bits
is accomplished by a \ac{DM}~\cite{schulte_constant_2016}.

Apart from closing the gap to the Shannon limit, \ac{PAS} also enables rate-adaptive transceiver designs~\cite[Sec.~VIII]{bocherer_bandwidth_2015} without 
changing the modcod. If a $2^m$-ASK constellation is used with a rate $c$ code, the \ac{SE} can be adjusted by the employed distribution $P_X$ and is given by
\begin{equation}
 R = \entr(X) - (1-c)m,\label{eq:se_pas}
\end{equation}
Hence, rate adaptation can be implemented by adjusting the \ac{DM} input parameter.

\begin{myremark}
 For uniform signaling, equation~\eqref{eq:se_pas} recovers $R=cm$.
\end{myremark}

\subsection{Numerical Comparisons with ATSC~3.0}

\begin{figure*}
 \footnotesize
 \subfloat[\SI{2.13}{\bpcu}]{\tikzsetnextfilename{coded_perf_2_13}\includegraphics{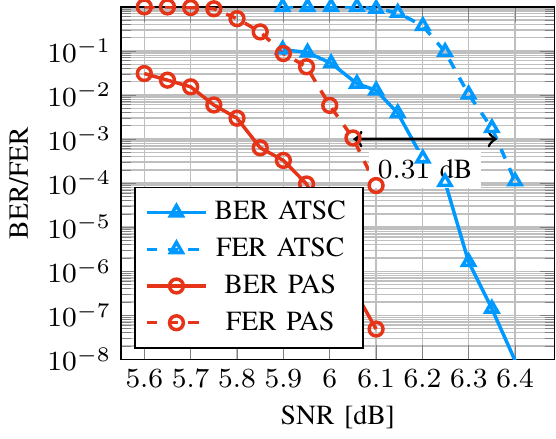}}
 \subfloat[\SI{3.20}{\bpcu}]{\tikzsetnextfilename{coded_perf_3_2}\includegraphics{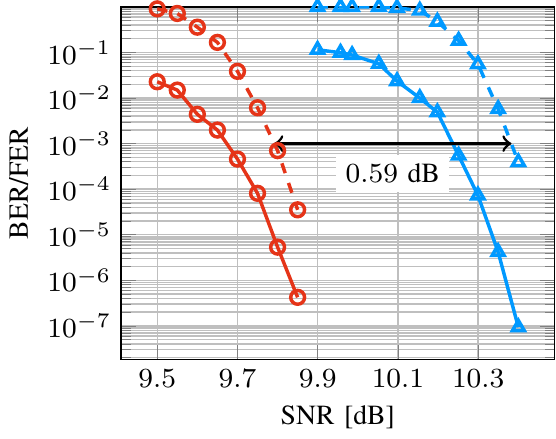}}
 \subfloat[\SI{5.33}{\bpcu}]{\tikzsetnextfilename{coded_perf_5_33}\includegraphics{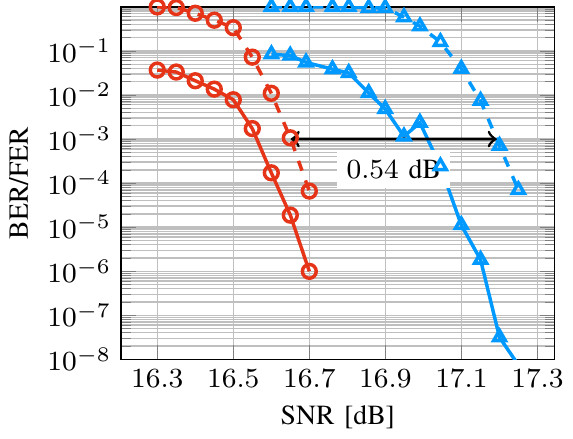}}
 \caption{Comparison of the coded performance of geometrically shaped ATSC~3.0 modcods with one single PAS scheme.}
 \label{fig:coded_performance}
\end{figure*}

ATSC~3.0 defines 6 constellations (QPSK, \{16, 64, 256, 1024, 4096\}-\acp{NUC}),
The smaller ones ($16, 64, 256$) are 2D-GS 2D-\acp{NUC}, 
whereas the the larger ones ($1024, 4096$) are 1D-GS 2D-NUCs.
The standard also defines \ac{LDPC} codes with blocklengths 16200 and 64800 bits for code rates from 2/15 to 13/15~\cite{kim_low-density_2016}, 
giving rise to 46 modcods for the long blocklengths and 29 modcods for the short blocklengths~\cite{michael_bit-interleaved_2016}.

\begin{figure}[h]
 \footnotesize
 \tikzsetnextfilename{atsc_perf}
 \includegraphics{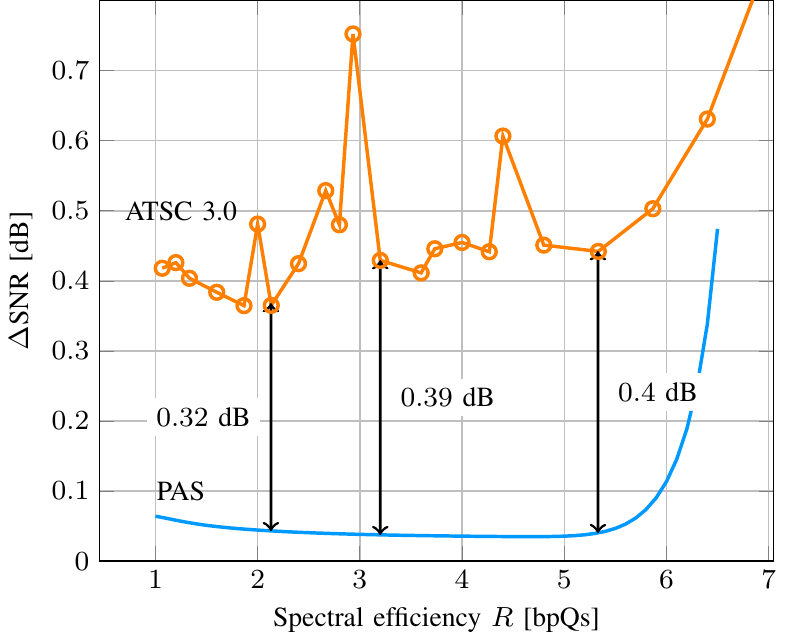}
 \caption{$\SNR$ gap to capacity for ATSC~3.0 \acp{AOP} comprising 2D-GS $\{16, 64, 256\}$ 2D-\acp{NUC} and allowed code rates compared to a single \ac{PAS} modcod of 256-QAM and a 5/6 code.}
 \label{fig:atsc_perf}
\end{figure}

For each modcod, the standard provides a constellation that has been designed to perform well with the associated
code. In Fig.~\ref{fig:atsc_perf}, we depict the \acp{AOP} of all mandatory ATSC~3.0 modcods~\cite[Table~6.12]{atsc30} involving the $\{16, 64, 256\}$-ary 2D-GS 2D-\acp{NUC}
by considering their gap to capacity. For each modcod, we calculate the required $\SNR_\text{req}$ to operate at an
\ac{SE} of $R=\log_2(M)\cdot c$ bits per channel use (\si{\bpcu}), i.e.,
$\SNR_\treq = R_\tbmd^{-1}(R)$.
For the \ac{PAS} case, we consider a 256-QAM constellation that is constructed by the Cartesian product of two equidistant 16-ASK constellations and a 5/6
rate code. Following \eqref{eq:se_pas}, the \ac{SE} is given as
\[
 R = \entr(X) - \left(1-\frac{5}{6}\right) \cdot 8 = \entr(X) - \frac{4}{3}
\]
which can be adjusted by modifying the entropy of $P_X$. To find the distribution $P_X^R$ that provides an \ac{SE} of $R$ \si{\bpcu},
we use the corresponding $\nu$ from the family of \ac{MB} distributions~\cite[Sec.~III-A]{steiner_design_robust_2016}. As before,
the required $\SNR$ is then given as $\SNR_\treq = \Rbmd^{-1}(P_X^R,R)$.
In both cases, the gap to capacity follows as $\Delta\SNR = \SNR_\treq - \sfC^{-1}(R)$.

We emphasize that only one single modcod is necessary for PAS to operate within the targeted \ac{SE} range of \SIrange{1.0}{5.33}{\bpcu} within \SI{0.06}{dB}.

In the following, we also compare the coded performance of a selection of modcods which are summarized in Table~\ref{tab:modcods_atsc}. This is
of interest to evaluate whether the calculated asymptotic gains of Fig.~\ref{fig:atsc_perf} derived from information theoretic quantities also translate into practice.

The employed rate 8/15 and 10/15 \ac{LDPC} codes for the ATSC~3.0 constellations are \ac{IRA} codes with blocklength \num{64800}. For
each constellation, a different interleaving and bit-mapping is employed according to the standard~\cite[Table~6.8]{atsc30}. \ac{PAS}
is operated with one single off-the-shelf 5/6 \ac{IRA} LDPC code from the DVB-S2 standard of the same blocklength with an optimized bit-mapper of~\cite[Sec. VII-B]{bocherer_bandwidth_2015}.
In both cases, 50 \ac{BP} iterations with full sum-product update rule at the check-nodes 
have been performed.

\begin{table}[h]
\caption{Summary of the considered modcods for \acp{SE} of \num{2.13}, \num{3.2} and \SI{5.33}{\bpcu}.}
\label{tab:modcods_atsc}
 \begin{tabular}{llcl}
  \toprule
      \ac{SE} [bpcu]  & Modcod & $R_{\tbmd}^{-1}(P_X,R)$ [\si{dB}] & Gap [\si{dB}]\\
  \midrule                                      
  \multirow{2}{*}{\num{2.13}} & PAS 256-QAM, 5/6 & \num{5.34} & \num{0.043}\\
                                      & ATSC 16-QAM, 8/15 & \num{5.66} & \num{0.37}\\
  \multirow{2}{*}{\num{3.20}} & PAS 256-QAM, 5/6 & \num{9.17} & \num{0.038}\\
                                      & ATSC 64-QAM, 8/15 & \num{9.56} & \num{0.43}\\
  \multirow{2}{*}{\num{5.33}} & PAS 256-QAM, 5/6 & \num{15.99} & \num{0.040}\\
                                      & ATSC 256-QAM, 10/15 & \num{16.38} & \num{0.44}\\                                      
  \bottomrule
 \end{tabular}
\end{table}

Looking at Fig.~\ref{fig:coded_performance}, we observe that the predicted performance gains can also
be observed in the coded results. For \acp{SE} of \SI{3.2}{\bpcu} and \SI{5.33}{\bpcu}, the gains in coded
performance even exceed the predicted ones (\SI{0.59}{dB} vs. \SI{0.39}{dB} and \SI{0.54}{dB} vs. \SI{0.4}{dB}).
Only for the lowest \ac{SE}, the gain is slightly smaller than expected (\SI{0.31}{dB} vs. \SI{0.32}{dB}).

\section{Conclusion}
\label{sec:conclusion}
This work presented a comprehensive comparison of \ac{PS} and \ac{GS}
for the \ac{AWGN} channel. We reviewed the underlying optimization problems and explained 
their mathematical properties. For the non-convex problem of \ac{GS}, a 
\ac{DE} approach was developed. The following achievable rate analysis 
shows that probabilistic shaping is able to close both the shaping and \ac{BMD} gap 
to approach \ac{AWGN} capacity, whereas a \SI{0.4}{dB} gap exits for \ac{GS}. 
Eventually, we compared the coded performance of different ATSC~3.0 \ac{GS} modcods 
to an equivalent setup with only one \ac{PS} modcod operated via \ac{PAS}. 
The predicted gains of the achievable rate analysis can also 
be observed in the coded scenario. This renders \ac{PAS} to become a viable candidate for 
rate-adaptive communication at the Shannon limit for future communication standards. 

\section*{Acknowledgement}
The authors would like to acknowledge the discussions with Dr.~David Gómez-Barquero and Manuel Fuentes Muela
concerning the ATSC~3.0 \ac{LDPC} codes. We also thank Dr.~Gianluigi Liva and Mustafa Cemil Coşkun for helpful comments
that greatly improved the presentation of this paper.



\begin{thebibliography}{10}
\providecommand{\url}[1]{#1}
\csname url@samestyle\endcsname
\providecommand{\newblock}{\relax}
\providecommand{\bibinfo}[2]{#2}
\providecommand{\BIBentrySTDinterwordspacing}{\spaceskip=0pt\relax}
\providecommand{\BIBentryALTinterwordstretchfactor}{4}
\providecommand{\BIBentryALTinterwordspacing}{\spaceskip=\fontdimen2\font plus
\BIBentryALTinterwordstretchfactor\fontdimen3\font minus
  \fontdimen4\font\relax}
\providecommand{\BIBforeignlanguage}[2]{{%
\expandafter\ifx\csname l@#1\endcsname\relax
\typeout{** WARNING: IEEEtran.bst: No hyphenation pattern has been}%
\typeout{** loaded for the language `#1'. Using the pattern for}%
\typeout{** the default language instead.}%
\else
\language=\csname l@#1\endcsname
\fi
#2}}
\providecommand{\BIBdecl}{\relax}
\BIBdecl

\bibitem{forney_efficient_1984}
G.~Forney, R.~Gallager, G.~Lang, F.~Longstaff, and S.~Qureshi, ``Efficient
  {{Modulation}} for {{Band-Limited Channels}},'' \emph{{IEEE} J. Sel. Areas
  Commun.}, vol.~2, no.~5, pp. 632--647, Sep. 1984.

\bibitem{bocherer_bandwidth_2015}
G.~B{\"o}cherer, F.~Steiner, and P.~Schulte, ``Bandwidth {{Efficient}} and
  {{Rate-Matched Low-Density Parity-Check Coded Modulation}},'' \emph{{IEEE}
  Trans. Commun.}, vol.~63, no.~12, pp. 4651--4665, Dec. 2015.

\bibitem{valenti_constellation_2012}
M.~Valenti and X.~Xiang, ``Constellation {{Shaping}} for {{Bit-Interleaved LDPC
  Coded APSK}},'' \emph{{IEEE} Trans. Commun.}, vol.~60, no.~10, pp.
  2960--2970, Oct. 2012.

\bibitem{sun_approaching_1993}
F.-W. Sun and H.~C.~A. {van Tilborg}, ``Approaching capacity by equiprobable
  signaling on the {{Gaussian}} channel,'' \emph{{IEEE} Trans. Inf. Theory},
  vol.~39, no.~5, pp. 1714--1716, Sep. 1993.

\bibitem{barsoum_constellation_2007}
M.~F. Barsoum, C.~Jones, and M.~Fitz, ``Constellation {{Design}} via {{Capacity
  Maximization}},'' in \emph{Proc. IEEE Int. Symp. Inf. Theory (ISIT)}, Jun.
  2007, pp. 1821--1825.

\bibitem{etsi2013dvbngh}
``Digital {{Video Broadcasting}} ({{DVB}});{{Next Generation}} broadcasting
  system to {{Handheld}},physical layer specification ({{DVB-NGH}}),'' no.
  A160, Nov. 2013.

\bibitem{gomez-barquero_dvb-ngh:_2014}
D.~G{\'o}mez-Barquero, C.~Douillard, P.~Moss, and V.~Mignone, ``{{DVB-NGH}}:
  {{The Next Generation}} of {{Digital Broadcast Services}} to {{Handheld
  Devices}},'' \emph{{IEEE} Trans. Broadcast.}, vol.~60, no.~2, pp. 246--257,
  Jun. 2014.

\bibitem{atsc30}
``{{ATSC Proposed Standard}}: {{Physical Layer Protocol}} ({{A}}/322),'' no.
  S32-230r56, Jun. 2016.

\bibitem{loghin_non-uniform_2016}
N.~S. Loghin, J.~Z{\"o}llner, B.~Mouhouche, D.~Ansorregui, J.~Kim, and S.~I.
  Park, ``Non-{{Uniform Constellations}} for {{ATSC}} 3.0,'' \emph{{IEEE}
  Trans. Broadcast.}, vol.~62, no.~1, pp. 197--203, Mar. 2016.

\bibitem{storn_differential_1997}
R.~Storn and K.~Price, ``\BIBforeignlanguage{en}{Differential {{Evolution}}
  \textendash{} {{A Simple}} and {{Efficient Heuristic}} for {{Global
  Optimization}} over {{Continuous Spaces}}},''
  \emph{\BIBforeignlanguage{en}{Journal of Global Optimization}}, vol.~11,
  no.~4, pp. 341--359, Dec. 1997.

\bibitem{bennatan_design_2006}
A.~Bennatan and D.~Burshtein, ``Design and analysis of nonbinary {{LDPC}} codes
  for arbitrary discrete-memoryless channels,'' \emph{{IEEE} Trans. Inf.
  Theory}, vol.~52, no.~2, pp. 549--583, Feb. 2006.

\bibitem{zehavi_8-psk_1992}
E.~Zehavi, ``8-{{PSK}} trellis codes for a {{Rayleigh}} channel,'' \emph{{IEEE}
  Trans. Commun.}, vol.~40, no.~5, pp. 873--884, May 1992.

\bibitem{caire_bit-interleaved_1998}
G.~Caire, G.~Taricco, and E.~Biglieri, ``Bit-interleaved coded modulation,''
  \emph{{IEEE} Trans. Inf. Theory}, vol.~44, no.~3, pp. 927--946, May 1998.

\bibitem{martinez_mismatched}
A.~Martinez, A.~{Guillen i Fabregas}, G.~Caire, and F.~Willems,
  ``Bit-{{Interleaved Coded Modulation Revisited}}: {{A Mismatched Decoding
  Perspective}},'' \emph{{IEEE} Trans. Inf. Theory}, vol.~55, no.~6, pp.
  2756--2765, Jun. 2009.

\bibitem{bocherer2014achievable}
G.~B{\"o}cherer, ``Achievable rates for shaped bit-metric decoding,''
  \emph{arXiv preprint 1410.8075v6}, 2016.

\bibitem{gray1953pulse}
F.~Gray, ``Pulse code communication,'' U. S. Patent 2\,632\,058, 1953.

\bibitem{zoellner_optimization_2013}
J.~Zoellner and N.~Loghin, ``Optimization of high-order non-uniform {{QAM}}
  constellations,'' in \emph{Proc. IEEE Int. Symp. Broadband Multim. Syst.
  Broadc. (BMSB)}, Jun. 2013, pp. 1--6.

\bibitem{kayhan_constellation_2012}
F.~Kayhan and G.~Montorsi, ``Constellation design for transmission over
  nonlinear satellite channels,'' in \emph{Proc. IEEE Global Telecommun. Conf.
  (GLOBECOM)}, Dec. 2012, pp. 3401--3406.

\bibitem{gallager1968}
R.~G. Gallager, \emph{Information {{Theory}} and {{Reliable
  Communication}}}.\hskip 1em plus 0.5em minus 0.4em\relax {John Wiley \& Sons,
  Inc.}, 1968.

\bibitem{blahut_capacity}
R.~E. Blahut, ``Computation of channel capacity and rate-distortion
  functions,'' \emph{{IEEE} Trans. Inf. Theory}, vol.~18, no.~4, pp. 460--473,
  1972.

\bibitem{huang_characterization_2005}
J.~Huang and S.~P. Meyn, ``Characterization and computation of optimal
  distributions for channel coding,'' \emph{{IEEE} Trans. Inf. Theory},
  vol.~51, no.~7, pp. 2336--2351, Jul. 2005.

\bibitem{kschischang_pasupathy_maxwell}
F.~Kschischang and S.~Pasupathy, ``Optimal nonuniform signaling for
  {{Gaussian}} channels,'' \emph{{IEEE} Trans. Inf. Theory}, vol.~39, no.~3,
  pp. 913--929, May 1993.

\bibitem{website_scc2016}
\BIBentryALTinterwordspacing
Collection of {O}ptimized {1D-GS} and {2D-GS} {NUCs}. [Online]. Available:
  \url{http://experimental-it.org/research/scc2017/}
\BIBentrySTDinterwordspacing

\bibitem{meric_approaching_2015}
H.~M{\'e}ric, ``Approaching the {{Gaussian Channel Capacity With APSK
  Constellations}},'' \emph{{IEEE} Commun. Lett.}, vol.~19, no.~7, pp.
  1125--1128, Jul. 2015.

\bibitem{wu_impact_2010}
Y.~Wu and S.~Verd{\'u}, ``The impact of constellation cardinality on
  {{Gaussian}} channel capacity,'' in \emph{Proc. Allerton Conf. Commun.,
  Contr., Comput.}, Sep. 2010, pp. 620--628.

\bibitem{schulte_constant_2016}
P.~Schulte and G.~B{\"o}cherer, ``Constant {{Composition Distribution
  Matching}},'' \emph{{IEEE} Trans. Inf. Theory}, vol.~62, no.~1, pp. 430--434,
  Jan. 2016.

\bibitem{kim_low-density_2016}
K.~J. Kim, S.~Myung, S.~I. Park, J.~Y. Lee, M.~Kan, Y.~Shinohara, J.~W. Shin,
  and J.~Kim, ``Low-{{Density Parity-Check Codes}} for {{ATSC}} 3.0,''
  \emph{{IEEE} Trans. Broadcast.}, vol.~62, no.~1, pp. 189--196, Mar. 2016.

\bibitem{michael_bit-interleaved_2016}
L.~Michael and D.~G{\'o}mez-Barquero, ``Bit-{{Interleaved Coded Modulation}}
  ({{BICM}}) for {{ATSC}} 3.0,'' \emph{{IEEE} Trans. Broadcast.}, vol.~62,
  no.~1, pp. 181--188, Mar. 2016.

\bibitem{steiner_design_robust_2016}
F.~Steiner and P.~Schulte, ``Design of robust, protograph based {LDPC} codes
  for {Rate-Adaptation} via probabilistic shaping,'' in \emph{Proc. Int. Symp.
  Turbo Codes and Iterative Inf. Process. (ISTC)}, Brest, France, Sep. 2016.

\end{thebibliography}
\end{document}